\documentclass[aps,prl,twocolumn,showpacs,superscriptaddress]{revtex4}
\usepackage{graphicx}

\begin{document}

\title{Phase Diagram of Heavy Fermion Metal CeCoIn$_5$}

\author{V.R. Shaginyan}
\email{vrshag@thd.pnpi.spb.ru}
\affiliation{Petersburg Nuclear
Physics Institute, Gatchina, 188300, Russia}
\affiliation{ CTSPS, Clark Atlanta University, Atlanta, Georgia
30314, USA}
\author{A.Z. Msezane}
\affiliation{ CTSPS, Clark Atlanta University, Atlanta, Georgia
30314, USA}
\author{V.A. Stephanovich}
\homepage{http://cs.uni.opole.pl/~stef}
\email{stef@math.uni.opole.pl}
\affiliation{Opole University,
Institute of Mathematics and Informatics, Opole, 45-052, Poland}
\author{E.V. Kirichenko}
\affiliation{Opole University,
Institute of Mathematics and Informatics, Opole, 45-052, Poland}

\begin{abstract}
We present a comprehensive analysis of the low temperature
experimental $H-T$ phase diagram of CeCoIn$_5$. The main universal
features of the diagram can be explained within the Fermi-liquid
theory provided that quasiparticles form so called
fermion-condensate state. We show that in this case the
fluctuations accompanying an ordinary quantum critical point are
strongly suppressed and cannot destroy the quasiparticles.
Analyzing the phase diagram and giving predictions, we demonstrate
that the electronic system of CeCoIn$_5$ provides a unique
opportunity to study the relationship between quasiparticles
properties and non-Fermi liquid behavior.
\end{abstract}


\maketitle

Although much theoretical efforts have been spared to understand
the non-Fermi liquid behavior (NFL) of heavy fermion (HF) metals
using the concept of quantum critical points, the problem is still
far from its complete understanding since the experimental systems
display serious discrepancies with the theoretical predictions
\cite{stew}. Common belief is that a quantum critical point (QCP)
is the point where a second order phase transition occurs at
temperature $T\to 0$, and where both thermal and quantum
fluctuations are present destroying quasiparticles and generating
a new regime around the point of instability between two stable
phases \cite{sac,col}. Recent experimental studies of the
CeCoIn$_5$  HF metal provide valuable information about the NFL
behavior near possible QCP due to its excellent tunability by a
pressure $P$ and/or a magnetic field $H$ \cite{pag,bi,ronn}. The
experimental studies have shown that besides a complicated $H-T$
phase diagram, the normal and superconducting properties around
the QCP exhibit various anomalies. One of them is power (in both
$T$ and $H$) variation of the resistivity and heat transport
\cite{pag,bi,ronn,mal,bauer}, inherent to  both NFL and Landau
Fermi liquid (LFL) regimes. The other one is a continuous magnetic
field evolution of a superconductive phase transition from the
second order to the first one \cite{izawa,bian}. Above anomalous
power laws can be hardly accounted for within scenarios based on
the QCP occurrence with quantum and thermal fluctuations. For
example, the divergence of the normal-state thermal expansion
coefficient, $\alpha/T$ is stronger than that in the 3D itinerant
spin-density-wave (SDW) theory, but weaker than that in the 2D SDW
picture \cite{pag1}. This brings the question of whether the
fluctuations are responsible for the observed behavior, and if
they are not, what kind of physics determines the above anomalies?
On the other hand, the direct observations of quasiparticle band
in CeIrIn$_5$ have been reported recently \cite{fujim}. However,
if the quasiparticles do exist, why they are not suppressed by the
fluctuations?

In this letter we show that these problems can be resolved within
LFL theory provided that quasiparticles form the so-called
fermion-condensate (FC) state \cite{ks,vol} emerging behind the
fermion condensation quantum phase transition (FCQPT) \cite{ams}.
We show that near FCQPT the fluctuations are strongly suppressed
while FC by itself is "protected" from above fluctuations by the
first order phase transition. We analyze the experimental $H-T$
phase diagram of CeCoIn$_5$ and show that its main universal
features can be well understood within the theory based on FCQPT.
We demonstrate that the electronic system of CeCoIn$_5$ can be
shifted from the ordered to disordered side of FCQPT by a magnetic
field, therefore giving a unique possibility to study the
relationship between quasiparticles and NFL behavior.

To study the low temperature universal features of HF metals, we
use the notion of HF liquid in order to avoid the complications
related to the crystalline anisotropy of solids. This is possible
since we consider the (universal) behavior related to the
power-law divergences of observables like the effective mass,
thermal expansion coefficient etc. These divergences are
determined by small (as compared to those from unit cell of a
corresponding reciprocal lattice) momenta transfer so that the
contribution from larger momenta can be safely ignored.

Let us consider HF liquid characterized by the effective mass
$M^*$. Upon applying the well-known equation, we can relate
$M^*$ to the bare electron mass $M$ \cite{pfit}
\begin{equation}\label{jk1}
\frac{M^*}{M}=\frac{1}{1-N_0F^1(p_F,p_F)/3}.
\end{equation}
Here $N_0$ is the density of states of a free electron gas, $p_F$
is Fermi momentum, and $F^1(p_F,p_F)$ is the $p$-wave component of
Landau interaction amplitude. Since LFL theory implies the
quasiparticle density in the form $x=p_F^3/3\pi^2$, we can rewrite
the amplitude as $F^1(p_F,p_F)=F^1(x)$. When at some $x=x_{\rm
FC}$, the $F^1(x)$ achieves some critical value, the denominator
in Eq. (\ref{jk1}) tends to zero so that the effective mass
diverges at $T=0$. Beyond the critical point $x_{\rm FC}$ the
denominator becomes negative making the effective mass negative.
To avoid physically meaningless states with $M^*<0$, the system
undergoes FCQPT with FC formation in the critical point $x=x_{\rm
FC}$. Therefore, behind the critical point $x_{\rm FC}$ the
quasiparticle spectrum is flat, $\varepsilon({\bf p})=\mu$, in
some region $p_i\leq p\leq p_f$ of momenta, while the
corresponding occupation number $n_0({\bf p})$ varies continuously
from 1 to 0, $0<n_0({\bf p})<1$ \cite{ks}. Here $\mu$ is a
chemical potential.

To investigate the FC state at $T=0$, we apply weak BCS-like
interaction \cite{bcs1} with the coupling constant $\lambda$ and
see what happens with the superconducting order parameter
$\kappa({\bf p})$ as $\lambda \to 0$. To do so, we express
$\kappa({\bf p})$ via the frequency integral from Gorkov function
$F^+$. To find latter function, we write the usual pair of
equations for functions $F^+$ and $G$ (see e.g. Ref.
\cite{lif_pit}). The solution of above equations yields
\begin{widetext}
\begin{eqnarray}
&&F^+(k,\omega )=-\frac{i\lambda <\kappa ({\bf p})>}{2\pi E_1({\bf
p})}\Biggl[ \frac{1}{\omega -E_1({\bf p})+i0}-\frac{1}{\omega
+E_1({\bf p})-i0}\Biggr],\nonumber \\
&&G(k,\omega )=\frac{1-n({\bf p})}{\omega -E_1({\bf
p})+i0}+\frac{n({\bf p})}{\omega +E_1({\bf p})-i0},\,\,\ E_1^2({\bf
p})=\left( \varepsilon({\bf p})-\mu \right) ^2+\Delta
^2,
\end{eqnarray}
\end{widetext}
where the superconducting gap,
\begin{equation}
\Delta ({\bf p})=-\lambda<\kappa({\bf p})>= -\lambda\int
\kappa({\bf p})\frac{d^3k}{(2\pi)^3},
\end{equation} with
\begin{equation}
\kappa({\bf p})=
\int_{-\infty }^{\infty }F^+({\bf p},\omega)\frac{d\omega
}{2\pi},
\end{equation}
is related to the dispersion $\varepsilon ({\bf p})$  as
\begin{equation}
\varepsilon ({\bf p})-\mu = \Delta ({\bf p})\frac{1-2v^2({\bf
p})}{2\kappa ({\bf p})}.
\end{equation}
Here $n({\bf p})=v^2({\bf p})$, the superconducting order parameter
$\kappa ({\bf p})=u({\bf p})v({\bf
p})=\sqrt{n({\bf p})(1-n({\bf p}))}$,
$u({\bf p})$ and $v({\bf p})$ are the
coefficients of corresponding Bogoliubov transformation, $u^2({\bf
p})+v^2({\bf p})=1$. Next we observe from Eqs. (3) and (5) that
when $\lambda\to0$ the dispersion $\varepsilon ({\bf p})$ becomes
flat, while Eq. (2) in the lowest order in $\lambda$ becomes
\begin{eqnarray}
&&F_0^+(k,\omega )=-i\sqrt{n_0({\bf p})(1-n_0({\bf p}))}
\left[ \frac{1}{\omega +i0}%
-\frac{1}{\omega -i0}\right] , \nonumber \\
&&G_0(k,\omega )=\frac{1-n_0({\bf p})}{\omega +i0}+\frac{n_0({\bf
p})}{\omega -i0},\,\,\,\,\,  \varepsilon({\bf p})-\mu =0.
\end{eqnarray}
To check this answer, we integrate $F_0^+$ over frequencies,
\begin{equation}\label{nfl7}
\int_{-\infty }^{\infty }F_0^+({\bf p},\omega )\frac{d\omega }{2\pi
}=\sqrt{n_0({\bf p})(1-n_0({\bf p}))},
\end{equation}
and conclude that $\kappa({\bf p})=\sqrt{n_0({\bf p})(1-n_0({\bf
p}))}$.

Since in our model the transition temperature $T_c\sim
\Delta\sim\lambda \to 0$, the order parameter $\kappa ({\bf p})$
vanishes at any finite temperature so that the quasiparticle
occupation number is given by Fermi-Dirac function which we
represent in the form
\begin{equation}\label{gol}
\varepsilon({\bf p},T) -\mu(T)=T\ln\frac{1-n({\bf
p},T)}{n({\bf p},T)}.
\end{equation}
Observing that at $T\to 0$ the distribution function satisfies the
inequality $0<n_0({\bf p},T)<1$ at $p_i\leq p\leq p_f$, we
conclude that both the entropy $S$ and the logarithm in the right
hand side of Eq. (\ref{gol}) are finite even at $T\to 0$. In this
case, the entropy $S_{\rm NFL}$ is related to the special solution
$n_0({\bf p})$ and contains the temperature independent term
$S_0=S_{\rm NFL}(T \to 0)$. For this special solution $n_0({\bf
p})$ the dispersion is flat, $\varepsilon({\bf p})=\mu$
\cite{zver}. Thus, the occupation number $n_0({\bf p})$ represents
the special solutions of both BCS and LFL equations determining
the NFL behavior of HF liquid. Namely, as it follows from Eq.
(\ref{nfl7}), contrary to conventional BSC case, the FC solutions
are characterized by infinitesimal value of superconducting gap,
$\Delta \to 0$, while the order parameter $\kappa_0({\bf p})$
remains finite and the entropy $S=0$. At the same time, in
contrast to the standard solutions of the LFL theory, the special
ones are characterized by the finite superconducting order
parameter $\kappa({\bf p})$ at $T=0$. At $T \to 0$ both the normal
state of the HF liquid with the finite entropy $S_{\rm NFL}$ and
the BCS state with $S=0$ coexist being separated by the first
order phase transition where the entropy undergoes a finite jump
$\delta S=S_0$. Due to the thermodynamic inequality,
\begin{equation}\label{etrm}
\delta Q\leq T\delta S,
\end{equation}
the heat $\delta Q$ of the transition is equal to zero making the
other thermodynamic functions continuous. Thus, both at the FCQPT
point and behind it there are no critical fluctuations
accompanying second order phase transitions and suppressing the
quasiparticles. As a result, the quasiparticles survive and define
the thermodynamic properties of the HF system.

On the basis of above special solutions peculiarities we can
explain the main universal properties of the $H-T$ phase diagram
of the HF metal CeCoIn$_5$ shown in Fig. 1. The latter substance
is a superconductor with $T_c=2.3$ K, while a field tuned QCP with
a critical field of $H_{c0}=5.1$ T coincides with $H_{c2}$, the
upper critical field where superconductivity vanishes
\cite{pag,bi}. We note that in some cases $H_{c0}=0$. For example,
CeRu$_2$Si$_2$ shows no magnetic ordering down to lowest
temperatures \cite{takah}. Therefore, in our simple HF model
$H_{c0}$ can be treated as fitting parameter. It follows from the
above consideration that $H_{c0}\simeq H_{c2}$ is an accidental
coincidence. Indeed, $H_{c2}$ is determined by  $\lambda$ which in
turn is given by the coupling of electrons with magnetic, phonon,
etc excitations rather than by $H_{c0}$. As a result, under
application of a pressure which influences differently $\lambda$
and $H_{c0}$, the above coincidence will be removed. This is in
agreement with the facts from Ref. \cite{ronn}.

\begin{figure}[!ht]
\begin{center}
\includegraphics[width=0.47\textwidth]{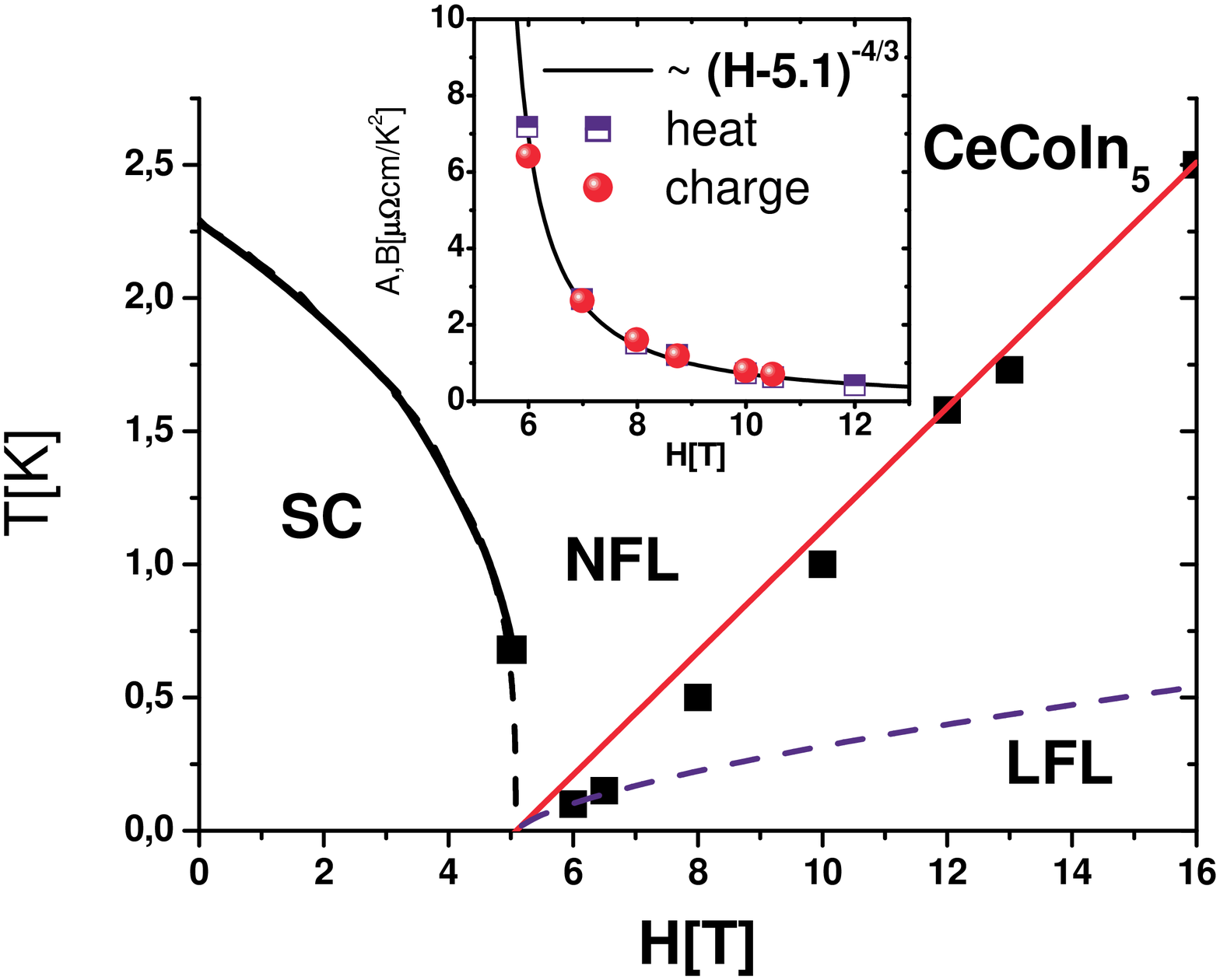}
\end{center}
\caption{$H-T$ phase diagram of CeCoIn$_5$. Superconducting-normal
phase boundary \cite{bian} is shown by the solid and dashed lines
with the solid square showing the point where the superconducting
phase transition changes from the second to the first order. The
solid squares (solid and dashed lines represent the calculated
dependence of the $T^*(H)$ as described in the text) obtained from
resistivity measurements denote the crossover boundaries between
the NFL and LFL regimes \cite{pag}. The inset shows the magnetic
field dependence of $T^2$ LFL coefficients of charge $A(H)\propto
H^{-4/3}$ and heat $B(H)\propto H^{-4/3}$ transport with
experimental data taken from Ref. \cite{pag}. } \label{Fig1}
\end{figure}

At relatively high temperatures, the superconducting-normal phase
transition in CeCoIn$_5$ shown by the solid line in Fig. 1 is of
the second order \cite{bian,izawa}. In that case, the entropy and
the other thermodynamic quantities are continuous at the
transition temperature $T_c(H)$. Since $H_{c2}\simeq H_{c0}$, upon
the application of magnetic field, the HF metal transits to its
NFL state down to lowest temperatures as seen from Fig. 1. As long
as the phase transition is of the second order, the entropy of SC
phase $S_{\rm SC}(T)$ coincides with the entropy $S_{\rm NFL}(T)$
of NFL state,
\begin{equation}\label{pq17}
S_{\rm SC}(T\to T_c(H))=S_{\rm NFL}(T\to T_c(H)).
\end{equation}
Since $S_{\rm SC}(T\to 0)\to 0$, Eq. (\ref{pq17}) cannot be
satisfied at sufficiently low temperatures due to the presence of
temperature-independent term $S_0$. Thus, in accordance with
experimental results \cite{bian,izawa}, the second order phase
transition converts to first order one below some temperature
$T_{0}(H)$. The prediction that the superconducting phase
transition may change its order had been made in the early 1960-s
\cite{maki}. This prediction is corroborated 
by our general analysis based 
on Eq. (\ref{pq17}). Namely, if the superconducting phase were
replaced by some other ordered phase separated from the NFL phase
by the second order phase transition at $H=0$, then at some
temperature $T_0(H)$ this phase transition should change its
order. In our case the NFL phase plays a role of disordered one.
But, as it follows from above consideration, the NFL phase has the
temperature independent entropy term $S_0$. Since in the ordered
phase the Nernst theorem ($S\to 0$ as $T\to 0$) should hold, we
come to conclusion that there is the entropy step (from $S_0$ to
zero) as $T\to 0$ while a system traverses the phase transition
line from ordered phase to NFL one. This means that this phase
transition should change its order at $T_0(H)$.
For example, the AFM phase transition with $T_N(H)$ (representing
the field dependence of N\'eel temperature) should become first
order at some finite temperature $T_{0}(H)$. Besides the jump in
the entropy, this first order transition, for example, should
result in a jump in the sample length, corresponding to a
divergence of $\alpha\propto -\partial S/\partial P$. Under
constant entropy (adiabatic) conditions, there should be a
temperature step as a magnetic field crosses the above phase
boundary due to inequality (\ref{etrm}). Indeed, the entropy jump
would release the heat, but since $S=const$ the heat is absorbed,
causing the temperature to decrease in order to keep the constant
entropy of the NFL state. Note that the minimal jump is given by
the temperature-independent term $S_0$, which can be quite large
so that the corresponding HF metal can be used as an effective
cooler at low temperatures.

The entropy $S_{\rm NFL}$ determines the anomalous behavior of
CeCoIn$_5$ in the NFL region of the phase diagram. The term
$S_0\sim(p_f-p_i)/p_F$ can be determined from the experimental
data on spin susceptibility (following Curie law) and the specific
heat jump $\Delta C$ at $T_c$ \cite{khzvya}. In HF metals like
CeCoIn$_5$ the normalized jump $\Delta C/C_n\simeq 4.5$ is
substantially higher than the ordinary BCS value \cite{petr},
where $C_n$ is the specific heat of a normal state. In the case of
FC, the specific heat jump is not proportional to $T_c$ and is
related to the fermion condensate parameter $\delta p_{\rm
FC}=(p_f-p_i)/p_F$, therefore the normalized jump $\Delta C/C_n$
can be large \cite{khzvya,ams1}. This estimation gives $\delta
p_{FC}\simeq 0.044$ \cite{khzvya}. The entropy $S_{\rm NFL}$
determines also both thermal expansion coefficient
$\alpha=-\partial S/\partial P$ and Gr\"uneisen ratio
$\Gamma=\alpha/C_n$ of  Fermi liquids with FC
\cite{zver,alp,khzvya}. Since the entropy has the temperature
independent part $S_0$, the thermal expansion coefficient
$\alpha\simeq-\partial S_0/\partial P$ becomes temperature
independent at low temperatures. Therefore, at $T \to 0$
$\alpha(T)\to const$ while the specific heat $C_n(T)\to 0$. As a
result, $\Gamma(T \to 0)$ diverges in coincidence with the facts
from Ref. \cite{pag1}.

Now we consider the LFL behavior tuned by a magnetic field $H\geq
H_{c0}$. Since the NFL behavior of CeCoIn$_5$ coincides with that
of YbRh$_2$(Si$_{0.95}$Ge$_{0.05}$)$_2$ and YbRh$_2$Si$_2$
\cite{cust,geg2} we would expect that  the LFL behavior of these
substances would also coincide. For example, in
YbRh$_2$(Si$_{0.95}$Ge$_{0.05}$)$_2$ the scattering coefficient
$A(H)$ in the resistivity $\rho=\rho_0+AT^2$, with $\rho_0$ being
the temperature independent part,
 diverges as $A(H)\propto (H-H_{c0})^{-1}$
\cite{cust} while in CeCoIn$_5$ it diverges as $A(H)\propto
(H-H_{c0})^c$ with the exponent $c\simeq -4/3$ \cite{pag,bi}. In
magnetic fields, the exponent $c=-1$ characterizes the function
$A(H)$ of HF liquid with FC \cite{shag}, while the exponent
$c=-4/3$ describes  the function $A(H)$ of HF liquid on the
disordered side of FC\cite{shag,clark}.

To understand this striking change in the behavior of CeCoIn$_5$,
we recall that FC has just appear in this substance since
$\delta p_{FC}=(p_f-p_i)/p_F \simeq 0.044\ll 1$. As soon as
magnetic field is sufficiently high, $H\geq H_{\rm cr}$, ($H_{\rm
cr}$ is a critical field destroying FC state), Zeeman splitting
$\delta p_F=(p_{F1}-p_{F2})/p_F$ of the two Fermi surfaces of HF
liquid exceeds the condensate parameter, $\delta p_F\geq p_{FC}$,
and the HF liquid with FC becomes LFL placed on the disordered
side near QCP. Here $p_{F1}$ and $p_{F2}$ are the Fermi momenta of
the two Fermi surfaces formed by the application of a magnetic
field. The splitting can be estimated as $p^2_F\delta
p_F/M^*(H)\sim H\mu_B$, where $\mu_B$ is Bohr magneton. Taking
into account that $A(H)\propto (M^*(H))^2$ we obtain
$(H_{cr}-H_{c0})/H_{c0}\sim (c_1\delta p_F)^{3}$. Our estimations
of the coefficient $c_1$ based on the experimental function $A(H)$
show that $c_1\sim 5$, and we obtain that reduced field
$(H_{cr}-H_{c0})/H_{c0}\sim (c_1\delta p_F)^3\simeq 0.02$. Thus,
we can safely suggest that the reduced field of $0.02$ is much
smaller then minimal reduced field $0.1$ where $A(H)$ measurements
have been carried out in Ref. \cite{pag,bi}. As a result, the
electronic system of CeCoIn$_5$ is placed on the disordered side
of FCQPT by the application of such high field and reveals
$A(H)\propto (H-H_{c0})^{-4/3}$. We can see from the inset to
Fig.1 that the coefficient $B(H)$ has the same critical field
dependence. Here $B(H)$ stands for the $T^2$-dependent
contribution to the thermal resistivity and related to $A(H)$ by a
field-independent factor, $A(H)/B(H)\simeq 0.47$, as it should be
in the case of ordinary metals \cite{pag}. We  conclude that the
LFL behavior of CeCoIn$_5$ corresponds to the LFL behavior of HF
liquid placed on the disordered side of FCQPT \cite{shag1}. At
decreasing field when $H<H_{cr}$, we predict that the exponent $c$
will change from $c=-4/3$ to $c=-1$.

At finite temperatures, the system remains in the LFL state, but
there exists a temperature $T^*(H)$ where the influence of FC is
recovered and the related  NFL behavior is restored. To calculate
the function $T^*(H)$, we observe that the effective mass $M^*$
cannot be changed at $T^*(H)$. Since at $T>T^*(H)$ the effective
mass $M^*(T)\propto 1/T$ \cite{noz} and at $T<T^*(H)$,
$M^*(H)\propto (H-H_{c0})^{-2/3}$, we have
$T^*(H)\propto (H-H_{c0})^{-2/3}.$
In high magnetic fields, there is a new crossover
line because the effective mass starts to depend on temperature as
$M^*(T)\propto T^{-2/3}$ \cite{clark,shag1} and $T^*(H)$ becomes
$T^*(H)\propto (H-H_{c0}).$
As it is seen from Fig. 1, the obtained results are in good
agreement with experimental facts. It is worth to note that the recovery of FC state can be observed
in measurements of a tunneling conductivity which is expected to
be noticeably asymmetrical with respect to the change of voltage
bias from $V$ to $-V$ in Fermi systems with FC \cite{tun}. This
asymmetry can be observed when the HF metal with FC is either
normal or superconducting because the asymmetry is indeed
determined by the violation of the so called particle-hole
symmetry, while this symmetry holds within the LFL theory. This
symmetry is not supported by FC occupation number $n_0({\bf p},T)$
since it does not evolve from the Fermi-Dirac distribution
function. Therefore, the tunneling conductivity is asymmetrical
when HF metal is superconducting, it does not change at $T=T_c$
and gradually vanishes at rising temperatures. We note also that
such behavior of the conductivity has recently been observed
experimentally in CeCoIn$_5$ \cite{park}. On the other hand, the
behavior of the conductivity can be different when the HF metal
transits from its LFL to NFL states. We predict that in the case
of CeCoIn$_5$ the conductivity being symmetrical in the LFL regime
becomes gradually asymmetrical reaching its maximum in the NFL
state at elevated temperatures when $T>T^*(H)$ and eventually
vanishes.

We thank P. Coleman for stimulating discussions. This work was
supported in part by RFBR, project No. 05-02-16085. The visit of
VRS to Clark Atlanta University has been upported by NSF through a
grant to CTSPS.

\end{document}